\documentclass[10pt, letter,twocolumn]{IEEEtran}

\usepackage[dvips]{color}
\usepackage{epsf}
\usepackage{times}
\usepackage{epsfig}
\usepackage{graphicx}
\usepackage{epstopdf}
\usepackage{algorithm}
\usepackage[noend]{algpseudocode}
\usepackage{amsmath}
\usepackage{amssymb}
\usepackage{amsxtra}
\usepackage{multirow}
\usepackage{mathtools}
\usepackage{cite}   
\usepackage{color}
\usepackage{bbm}
\usepackage{subfigure}
\usepackage{nomencl}
\usepackage{verbatim}
\usepackage{multirow}
\usepackage[hyphens]{url}
\usepackage{hyperref}
\newlength{\continueindent}
\usepackage{etoolbox}
\makeatletter
\newcommand*{\ALG@customparshape}{\parshape 2 \leftmargin \linewidth \dimexpr\ALG@tlm+\continueindent\relax \dimexpr\linewidth+\leftmargin-\ALG@tlm-\continueindent\relax}
\apptocmd{\ALG@beginblock}{\ALG@customparshape}{}{\errmessage{failed to patch}}
\renewcommand{\ALG@beginalgorithmic}{\small}
\makeatother

\title{Elevated LiDAR Placement under Energy and Throughput Capacity Constraints} 

\author{
\IEEEauthorblockN{\large Michael C. Lucic, Hakim Ghazzai, and Yehia Massoud}\\
\IEEEauthorblockA{\small School of Systems and Enterprises -- Stevens Institute of Technology, Hoboken, NJ, USA} 
{\thanks {\hrule
\vspace{0.1cm} \indent This paper is accepted for publication in the IEEE 63rd Midwest Symposium on Circuits \& Systems (MWSCAS 2020), Springfield, MA, USA, Aug. 2020.  

\textcopyright~2020 IEEE.  Personal use of this material is permitted.  Permission from IEEE must be obtained for all other uses, in any current or future media, including reprinting/republishing this material for advertising or promotional purposes, creating new collective works, for resale or redistribution to servers or lists, or reuse of any copyrighted component of this work in other works.
}}
}

\begin{document}
\maketitle
\thispagestyle{empty}
\begin{abstract}
\boldmath
Elevated LiDAR (ELiD) has the potential to hasten the deployment of Autonomous Vehicles (AV), as ELiD can reduce energy expenditures associated with AVs, and can also be utilized for other intelligent Transportation Systems applications such  as  urban  3D  mapping.  In  this  paper,  we  address  the  need for  a  planning  framework  in  order  for  ITS  operators  to  have an effective tool for determining what resources are required to achieve a desired level of coverage of urban roadways. To this end, we  develop  a  mixed-integer  nonlinear  constrained  optimization problem,  with  the  aim  of  maximizing  effective  area  coverage of  a  roadway,  while  satisfying  energy  and  throughput  capacity constraints.  Due  to  the  non-linearity  of  the  problem,  we  utilize Particle Swarm Optimization (PSO) to solve the problem. After demonstrating its effectiveness in finding a solution for a realistic scenario,  we  perform  a  sensitivity  analysis  to  test  the  model's general  ability. 
\end{abstract}

\section{Introduction}
\label{intro}

Autonomous Vehicles (AV) provide the potential to dramatically improve the safety and operation of urban roadways. AV operations are dependent on the fusion of data streams from three core sensing technologies: cameras, radar, and Light Detection and Ranging (LiDAR), in order for the vehicle to effectively map its surroundings for navigation purposes~\cite{6179503, 7747236}. In addition to their use as the main sensor for 3D mapping surroundings, LiDARs are used in a variety of applications, such as urban 3D mapping/infrastructure health monitoring~\cite{6516106}.

Due to the fine granularity in which they are capable of mapping their surroundings, LiDARs provide a very robust set of data for AVs to construct a detailed map of their surroundings. Nevertheless, this wealth of data comes with a series of costs. First, LiDARs are expensive; a Velodyne LiDAR that is commonly used in AV applications costs USD 8k per unit~\cite{8952784}. Second, the field-of-view (FoV) perspective of a LiDAR in a car limits the in the amount of data it can capture from its surroundings. Third, AV LiDARs and their complementary on-board systems are a huge drain on the AV's energy reserves. The data generated by LiDARs are similar to that of image data, so power-hungry GPUs are oftentimes utilized to accelerate LiDAR data processing. As a result of this, AVs have 10\% worse fuel efficiency than their conventional vehicle counterparts.

\begin{figure}[t!]
\centering
\vspace{0.0cm}
\includegraphics[width=7cm]{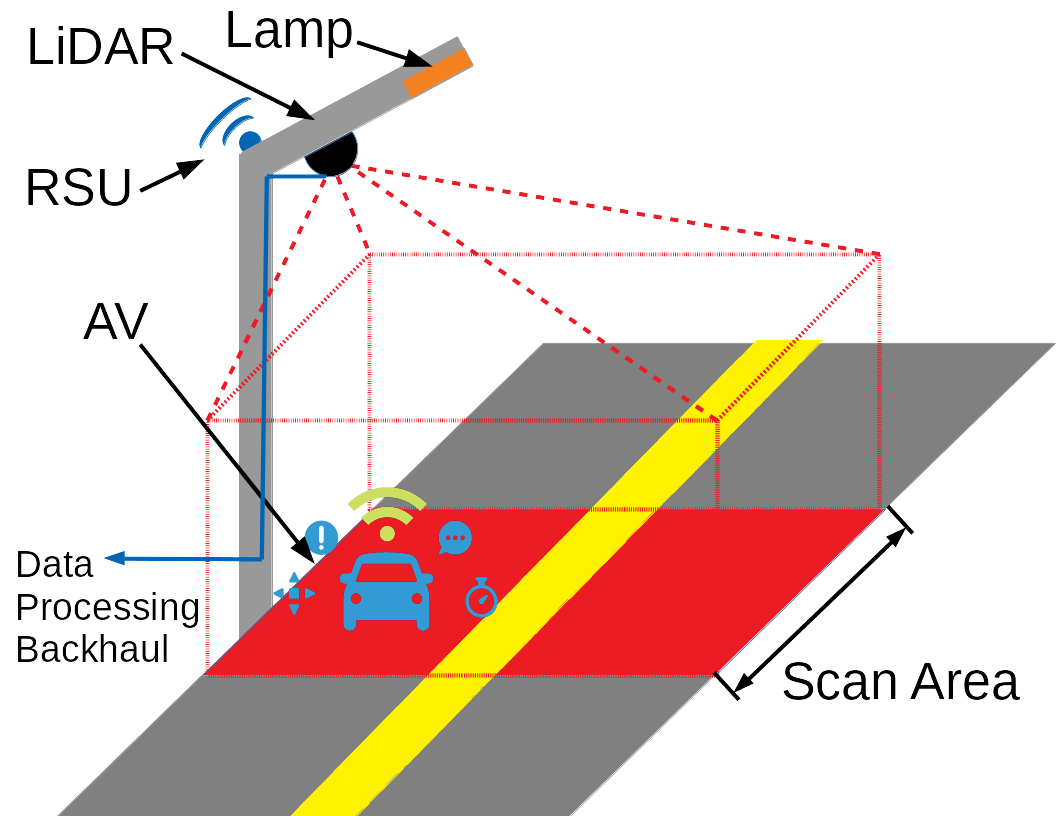}
\caption{Elevated LiDAR effective scan area (Isometric View)}
\label{scanAreaIsometric}
\end{figure}

\begin{figure*}[t!]
\centering
\begin{tabular}{cc}
\vspace{0.0cm}
\includegraphics[width=7.5cm]{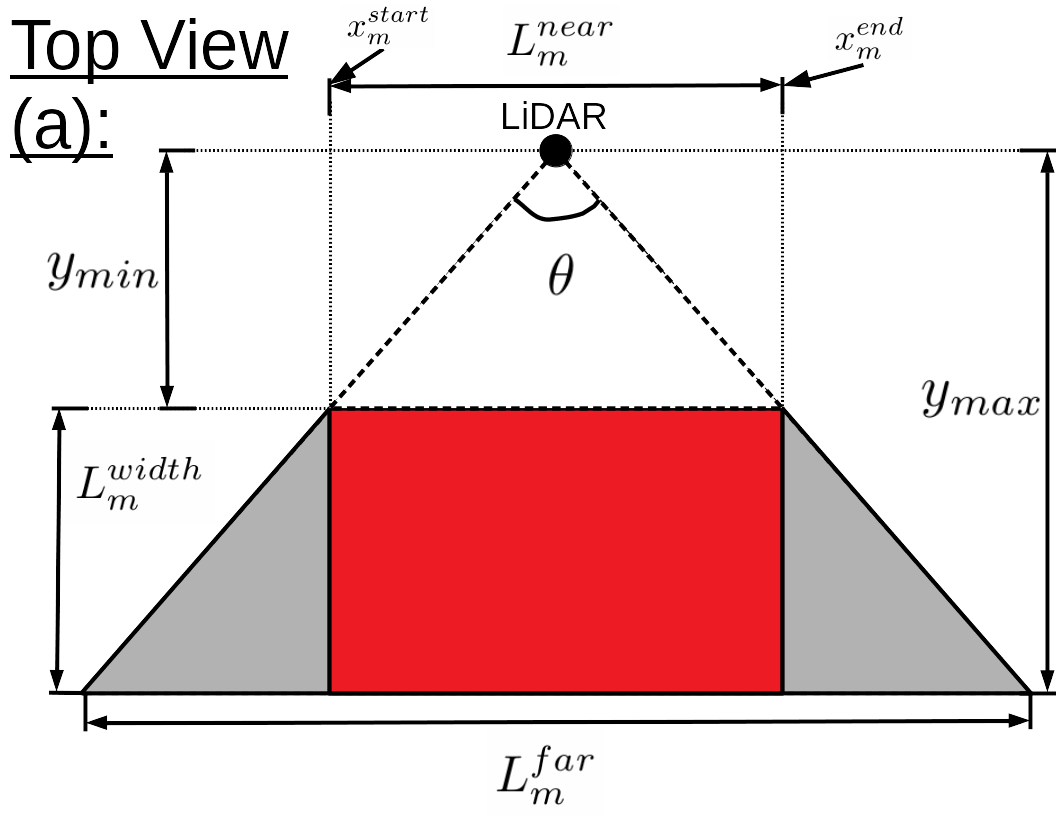}&
\includegraphics[width=7.5cm]{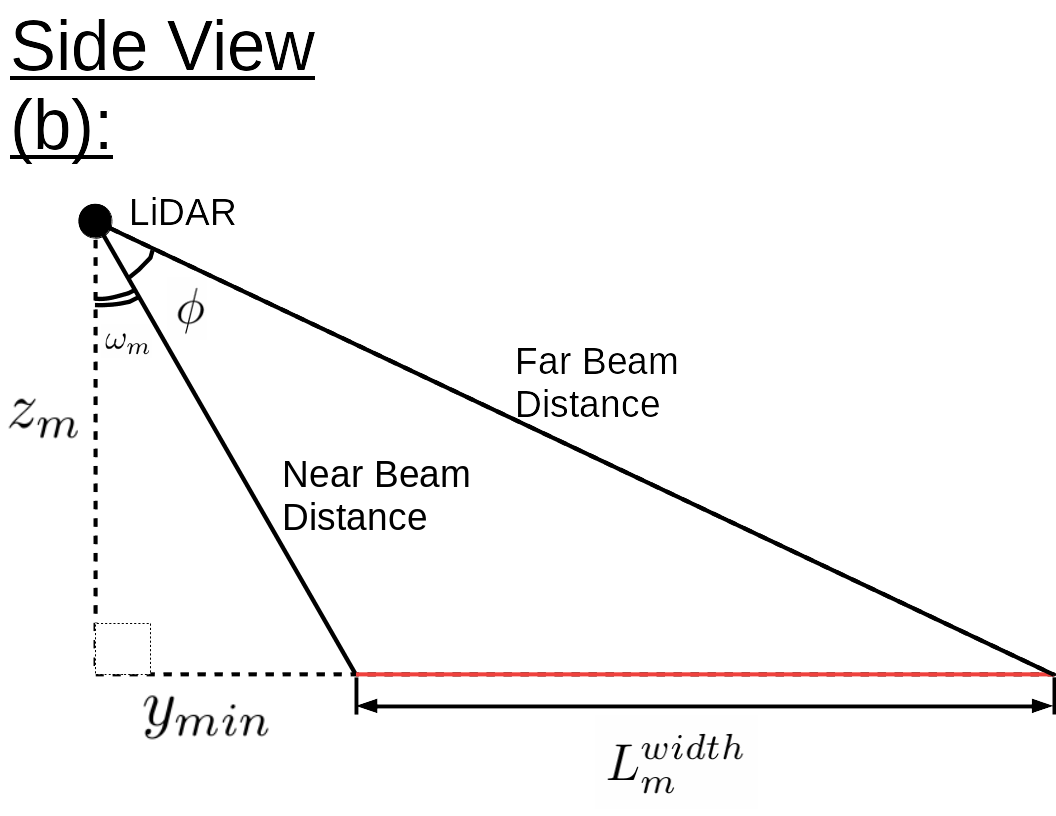}\\
\includegraphics[width=5cm]{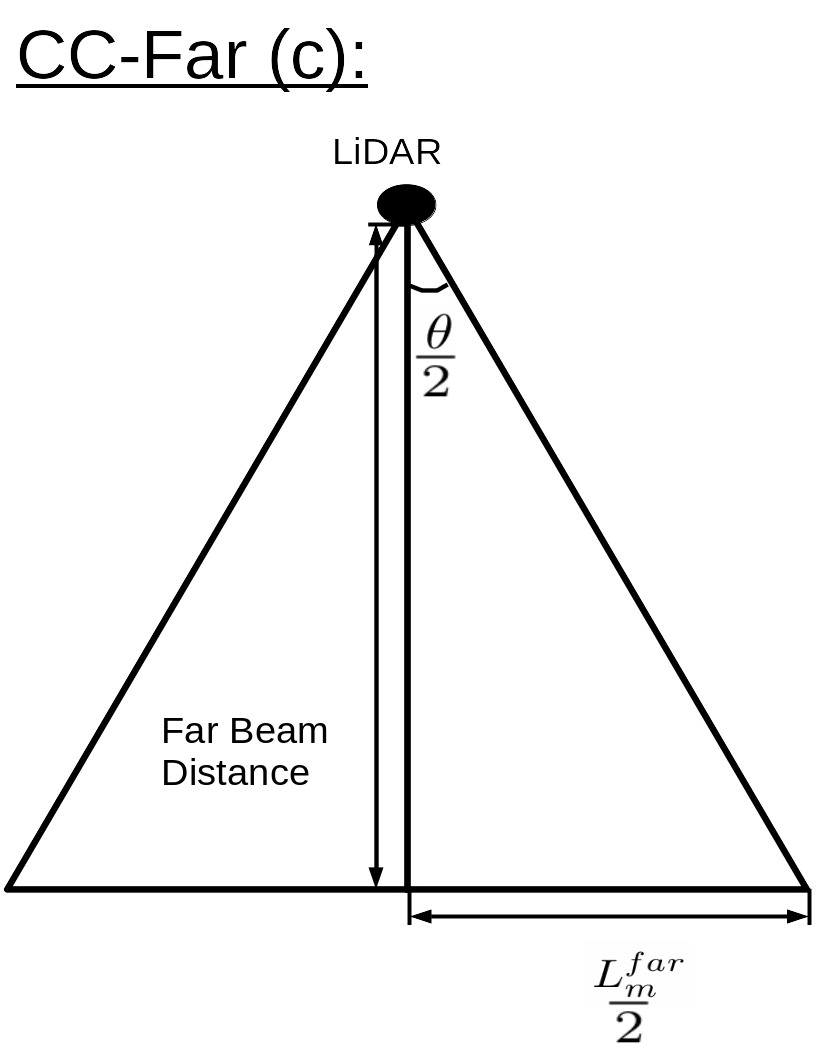}&
\includegraphics[width=5cm]{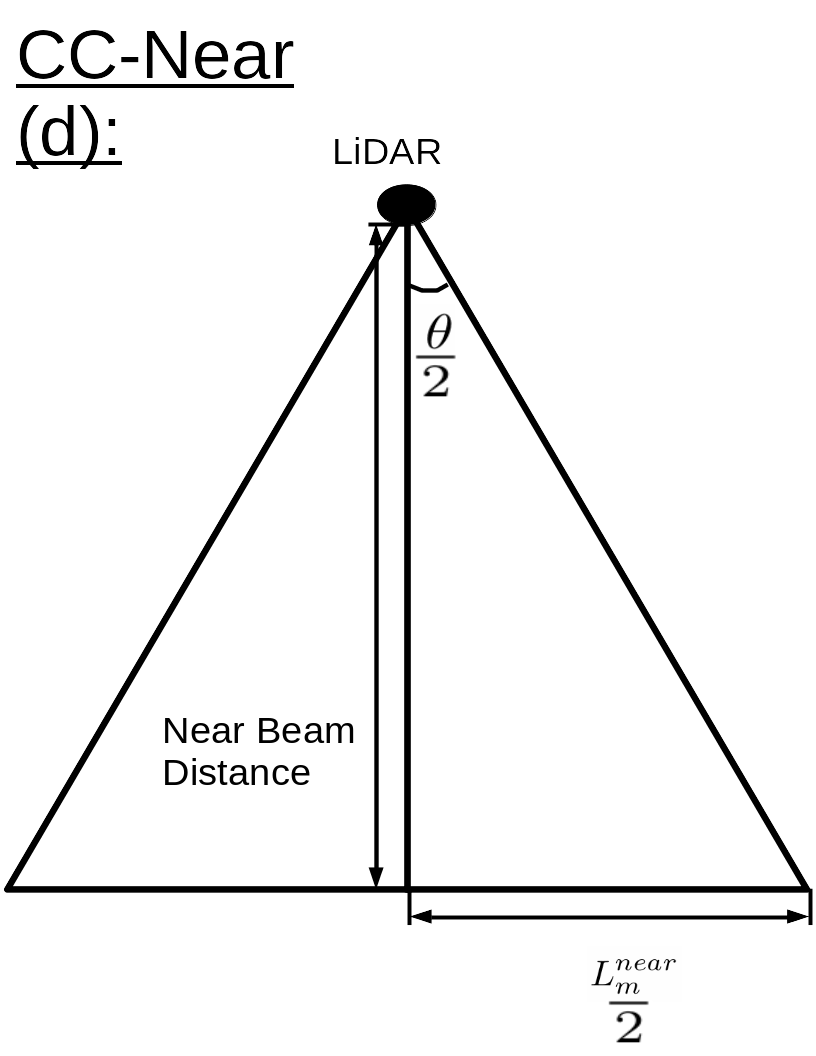} 
\end{tabular}
\caption{Elevated LiDAR effective scan area. (a): top down view. (b): side view - from perspective of car on the road. Other Two: cross section view, perpendicular to the roadway, facing toward (c), and away from (d) the ELiD.}
\label{scanAreaCC}
\end{figure*}

To mitigate the issues associated with the use of LiDAR in autonomous vehicles, a novel elevated LiDAR (ELiD) system has been proposed~\cite{8746507, 8952784, lucic2020latency}. ELiD moves LiDARs to elevated positions such as getting mounted onto the side of buildings or on street lamps (see Fig.~\ref{scanAreaIsometric}) along with wireless roadside units (RSUs) that can communicate with vehicles driving by. This allows for AVs to reduce their reliance on on-board LiDAR, making them more efficient, while improving the effective roadway scanning coverage with fewer units. The authors of~\cite{8746507} have proposed various challenges that must be addressed in order to make ELiD viable for deployment, including the development of a backhaul network. The authors of~\cite{8952784} developed a problem formulation with the aim of minimizing costs associated with allocating ELiD processing tasks to virtual machine instances in a hybrid edge/cloud network. In another approach, the authors of~\cite{lucic2020latency} aim to minimize latency associated with allocating tasks between the edge and the cloud. Other approaches to optimized task allocation in mobile edge computing backhaul networks have been explored as well~\cite{8490653, 7749210, 7222400, 7090242}. Configuration of roadside units (RSUs) needed to connect AVs to the ELiD's in a way to maximize effective area coverage while remaining within an amortized budget has been explored by the authors of~\cite{9000789}.

In this paper, we develop a placement method for multiple ELiD's along a roadway to maximize effective coverage area of an urban roadway, while taking into account the energy constraints and data throughput caps. To this end, we model the problem as a novel non-linear mixed-integer programming problem while considering ELiD lamps coverage overlaps. Due to the non-linearity of the objective function and its complexity, we propose to solve the constrained problem using the Particle Swarm Optimization (PSO) algorithm. The PSO solver is capable of finding an effective ELiD configuration for a realistic setting. 


\section{System Model}
\label{sysmod}
We consider a set of potential ELiD's $m \in \mathcal{M}$, where each ELiD lamp has identical technical specifications, such as their horizontal field-of-view (FoV) angle $\theta$, vertical FoV angle $\phi$, and scan rate $f_{scan}$. The ELiD's are mounted on buildings or lamp posts that are above and to the side of the road, such that the coverage surface of the lamp can be modeled as a trapezoid. Fig.~\ref{scanAreaIsometric} provides an isometric view of how each lamp covers the roadway. The area of roadway covered by the elevated LiDAR is affected by its height $z_m \in [z_{min}, z_{max}]$, vertical rotation angle $\omega_m \in [0, \frac{\pi}{2} - \phi)$, and horizontal position $x_m \in [0, D_{road}]$. These ELiDs are covering an urban roadway with a length of $D_{road}$ meters by a width of $y_{max} - y_{min}$ meters, where $y_{max}$ is the distance from the ELiD to the far side of the road, and $y_{min}$ is the distance from the ELiD to the near side of the road. Since certain sections of the roadway may have more activity than others (i.e. an intersection), the roadway can be split into $A$ sectors, where each sector $a \in \{1, \ldots, A\}$ has a relevance score $\Delta_a \in [0, 1]$, where a higher score corresponds to a higher amount of traffic activity. Each sector has an end location $Q_a$, such that $Q_a \in [0, D_{road}]$. Each sector shares the same width as the roadway (i.e. $y_{max} - y_{min}$). Fig.~\ref{scanAreaCC} provides multiple viewpoints of how the ELiD coverage zone projects onto the roadway.

From the visualization of the lamp coverage area, we notice that the coverage area can be modeled as a trapezoidal area $A^{Total}_m$, where the near base length $L^{near}_m$, far base length $L^{far}_m$, and coverage width $L^{width}_m$ that comprise the dimensions of the coverage trapezoid. In order to consider the effective coverage of the ELiD when installed in conjunction with other ELiD units, it is worthwhile to consider the effective area $A^{Rectangle}_m$, which is just the smaller rectangular area within the coverage trapezoid. Geometric representations of this are visualized in Fig.~\ref{scanAreaCC}. The expressions for the coverage areas and dimensions can be expressed as follows:
\begin{subequations}
\small
\begin{align}
&A^{Total}_m = \frac{(L^{near}_m + L^{far}_m) \times L^{width}_m}{2}, &\forall m \in \mathcal{M}\\
&A^{Rectangle}_m = L^{near}_m \times L^{width}_m, &\forall m \in \mathcal{M} \\
&L^{near}_m = 2z_m\sec{(\omega_m)}\tan{\left(\frac{\theta}{2}\right)}, &\forall m \in \mathcal{M}\\
&L^{far}_m = L^{near}_m \times \frac{\sec{(\omega_m + \phi)}}{\sec{(\omega_m})}, &\forall m \in \mathcal{M}\\
&L^{width}_m = z_m[\tan{(\omega_m + \phi)} + \tan{(\omega_m)}], &\forall m \in \mathcal{M}.
\end{align}
\end{subequations}
The expressions for $L^{near}_m$, $L^{far}_m$, and $L^{width}_m$ are derived from the geometric relationships that are visualized in Fig.~\ref{scanAreaCC}.

The data structure commonly utilized to store data captured by an ELiD scan is an octree \cite{Kumar2012}. In an octree, the scan sector is divided into 8 sub-sectors if an object is detected in the scan area, which is then repeated recursively $d$ times. The scan area in a sub-sector is expanded only if an object is detected, thus a scan area may have a very large set of possible scan permutations. Based on this, in a worst-case scenario where the maximum number of subdivisions are required to scan an area, each scan may generate $G_{cov}$ bytes/m$^3$ of data, which can be expressed as:
\begin{equation}
\small
G_{cov} = 8^{d - 2} + 12,
\end{equation}
where the $12$ bytes represent a 3-tuple of 32-bit floats that is needed to define the origin point of the scan. Based on the assumption that the ELiDs generate octree point set clouds, total amount of data generated by the ELiD $D_m$ can be approximated as follows:
\begin{equation}
\small
D_m = H_{cov} \times A^{Total}_m \times G_{cov} \times f_{scan}^{-1},
\end{equation}
where $H_{cov}$ is the height of the coverage area detection zone.

We can then approximate the energy consumption of each ELiD $E_m$ based on the amount of data generated and its scan frequency. This approximate energy consumption is expressed as follows:
\begin{equation}
\small
E_m = (P_{comm} \times D_m \times R_{comm}^{-1}) + (P_{rad} \times f_{scan}^{-1}),
\end{equation}
where $R_{comm}$ is the data upload transmission rate in bytes/s from the ELiD, $P_{comm}$ is the power in Watts required to operate the equipment that communicates the ELiD data, and $P_{rad}$ is the laser scan power for the ELiD.



\section{Problem Formulation}
\label{formulation}

In this section, we formulate a constrained mixed-integer non-linear programming (MINLP) problem aiming to maximize the effective area coverage of the deployed ELiDs deployed, while satisfying energy and throughput constraints. 

\subsection{Objective Function}
\label{objFunc}
Our objective is to install a series of ELiDs that can effectively scan the roadway; maximizing their effective coverage. When considering the effective area coverage, overlaps in the coverage regions of the ELiD's must be taken into account - an area covered by two ELiD's can only be considered as covered once.

To aid the construction of the objective function, we assume that a single ELiD coverage area can be modeled like the red square illustrated in Fig.~\ref{scanAreaCC}(a) (i.e. $A^{Rectangle}_m$). For each ELiD $m$, the start ($x^{start}_m$) and end ($x^{end}_m$) points of the coverage length along the roadway (see Fig~\ref{scanAreaCC}(a)) can be defined as:
\begin{subequations}
\small
\begin{align}
&x^{start}_m = \max{(x_m - L^{near}_m, 0)}\\
&x^{end}_m = \min{(x_m + L^{near}_m, D_{road})}.
\end{align}
\end{subequations}
The ELiDs may be configured in a way that leads to their coverage zones lying outside of the boundaries of the rectangle that defines the roadway - the max and min restrictions ignore this extraneous coverage outside of the area of interest - it is waste in the scope of this problem.

These $x^{start}_m$ and $x^{end}_m$ points can be defined as "Points of Interest" (PoIs), where some boundaries in the coverage areas occur. We therefore define a set of PoI $\Psi \in \mathbb{R}^{2|\mathcal{M}| + 2}$, where $\psi_j \in \Psi = \{0, x^{start}_m, \forall m \in \mathcal{M}, x^{end}_m, \forall m \in \mathcal{M}, D_{road}\}$, and $\Psi$ is sorted in ascending order. The rectangular area of the coverage sub-rectangles $A^{sub}_j$ of the roadway partitioned by the ELiD PoI $\Psi$ can then be expressed as:
\begin{subequations}
\small
\begin{align}
&A^{sub}_j = L^{sub}_j \times W^{sub}_j: L^{sub}_j = \psi_{j} - \psi_{j-1}, \\
&W^{sub}_j = \max\limits_{m \in \mathcal{M}}{\{\kappa_{jm}\epsilon_m(\min{\{L^{width}_m, y_{max} - y_{min}\})}}\}
\end{align}
\end{subequations}
where $\epsilon_m$ is a binary decision variable that corresponds to whether an ELiD is ($\epsilon_m = 1$) or is not placed ($\epsilon_m = 0$), and $\kappa_{jm}$ is a binary indicator variable, which is equal to 1 if an ELiD $m$ is covering the sub-rectangle $j$, or 0 otherwise.

In addition, we must consider the weighted relevance score $\Gamma_j$ in the situation that the sub-rectangles split the relevance sectors $a$. It can be expressed as follows:
\begin{equation}
\small
\Gamma_j = \frac{1}{L^{sub}_j}\sum_{a = 1}^{A}{\mu_{aj}\Delta_a(\min{(\psi_{j}, Q_{a})} - \max{(\psi_{j-1},  Q_{a-1})})}
\end{equation}
where $\mu_{aj}$ is a binary indicator variable, which is equal to 1 if a relevance sector $a$ is located within the sub-rectangle $j$, or 0 otherwise. With consideration of the aforementioned derived expressions, we can express the effective coverage ratio $A^{eff}_{cov}$ (i.e. the objective function to maximize) as follows:
\begin{equation}
\small
A^{eff}_{cov} = \frac{1}{\eta(y_{max} - y_{min})D_{road}}\sum_{j \in 2, \ldots, 2|\mathcal{M}| + 2}\Gamma_j A^{sub}_j,
\label{objective}
\end{equation}
where $\eta$ is the fraction of the roadway that must be covered (i.e. if $\eta = 0.99$, then 99\% of the defined roadway must be covered by the array of lamps).

\subsection{Constraints}
\label{constraints}
\subsubsection{Throughput Constraint}
\label{bandConstr}
The total rate that data are generated ($D_{m}$) by the ELiDs $m \in \mathcal{M}$ should not exceed the network throughput limit ($\bar{B}$) of the fiber links, as defined in the following:
\begin{equation}
\small
\frac{\sum_{m \in \mathcal{M}}\epsilon_mD_m}{\bar{B}} - 1 \leq 0.
\label{band}
\end{equation}

\subsubsection{Energy Constraints}
\label{energyConstr}
The energy consumed by the ELiD in operation should not exceed safety limits ($\bar{E}$) as given below:
\begin{equation}
\small
\frac{\epsilon_mE_m}{\bar{E}} - 1 \leq 0, \forall m \in \mathcal{M}.
\label{energy}
\end{equation}

\subsubsection{ELiD Orientation Constraints}
\label{yDistConstr}
The ELiD's vertical rotation must make the lowest ELiD laser beam point to the closest point of the road (i.e. $y_{min}$) as indicated by the following constraint:
\begin{equation}
\small
\omega_{m} = \arctan{\left(\frac{y_{min}}{z_{m}}\right)}, \forall m \in \mathcal{M}.
\label{width}
\end{equation}

\subsection{MINLP Formulation}
\label{MINLPFormulation}
The MINLP to maximize effective coverage of the roadway by the ELiDs can be expressed as follows:
\begin{align*}
\small
(\mathcal{P})&\underset{{x_{m}\in [0, D_{road}],\atop z_{m}\in [z_{min}, z_{max}], \epsilon_{m}\in \{0, 1\}}}{\text{minimize}}\quad-A^{eff}_{cov} + \lambda \sum_{m \in \mathcal{M}} \epsilon_m\\
&\text{s.t. }\quad \text{\small Constraints \eqref{band} - \eqref{width}}.
\end{align*}
The second term in the objective function is added to limit the number of ELiDs placed to minimize the amount of overlapped coverage where $\lambda$ is a regularization parameter.



The problem $(\mathcal{P})$ can be expressed as a general constrained non-linear programming problem, where the objective function is $f(\boldsymbol{X}) = -A^{eff}_{cov} + \lambda \sum_{m \in \mathcal{M}} \epsilon_m$, the inequality constraints can be expressed as $h_{im}(x_m, z_m, \epsilon_m) \leq 0$ for constraints \eqref{band} and \eqref{energy}, and all instances of $\omega_m$ can be replaced with $\arctan{\left(\frac{y_{min}}{z_{m}}\right)}$. Note that $x_m, z_m, \epsilon_m \in \boldsymbol{X}, \forall m \in \mathcal{M}$ (i.e. $\boldsymbol{X}$ is used for notation compactness).

\subsection{Particle Swarm Optimization Heuristic Development}
\label{PSO}
Since the objective function and constraint~\eqref{width} are highly nonlinear, we must utilize approximation methods to solve the problem. Indeed, it is very hard to find an optimal solution for these kinds of problems involving binary and continuous decisions variables and non-linear objective functions and constraints. One well-studied optimization technique for non-linear optimization is the Particle Swarm Optimization (PSO) for continuous search spaces~\cite{488968}. Since we have a mixed-integer problem, we utilize binary PSO (BPSO)~\cite{kennedy1997discrete} in parallel.

In classic PSO, the velocity ($v_{ms}^{t}$) and particle ($x_{ms}^{t}$) update rules for a continuous variable $x$ (this would be used for $x_m$ and $z_m$) are formulated as:
\begin{subequations}
\begin{align}
\small
\small v_{ms}^{t+1} &= \alpha v_{ms}^{t} + \beta_{p} r_{p} (p_{ms} - x_{ms}^{t}) + \beta_{g} r_{g} (g_{m} - x_{ms}^{t}),\label{velEq}\\
x_{ms}^{t+1} &= x_{ms}^{t} + v_{ms}^{t+1},
\end{align}
\end{subequations}
where $t$ is the current iteration, $s \in S$ is the particle index (of $|S|$ particles), $p_{ms}$ and $g_{m}$ are the local and global best for a particle. respectively, $\alpha$ is the inertia, $\beta_{p}$ and $\beta_{g}$ are the local and global best weights, respectively, and $r_{p}$ and $r_{g}$ are $U(0, 1)$ random values. Since $\epsilon_{m}$ is binary, after updating its velocity with equation~(\ref{velEq}), we utilize the BPSO update rules for this variable instead:
\begin{subequations}
\begin{align}
\small
\sigma_{ms}^{t+1} &= (1 + \exp{\{v_{ms}^{t+1}\}})^{-1}\\
\epsilon_{ms}^{t+1} &= \begin{cases}1,& \text{if } r_{b} < \sigma_{ms}^{t+1} \\ 0,& \text{otherwise}  \end{cases},
\end{align}
\end{subequations}
where $r_b \sim U(0, 1)$ is a random value. The PSO procedure terminates after $t_{max}$ iterations or there is no improvement larger than $\xi$ in in the fitness function for $\frac{1}{10}t_{max}$ consecutive iterations.



PSO is typically applied to solve unconstrained optimization problems. To this end, the constraints can be transformed into exterior penalty functions which can be incorporated into the objective function -- effectively transforming the problem into an unconstrained non-linear optimization problem. These penalty functions can be expressed as follows:
\begin{equation}
\small
P_{im}(x_m, z_m, \epsilon_{m}) = \rho_{im}\max\{h_{im}(x_m, z_m, \epsilon_{m}), 0\}^2,
\end{equation}
where $\rho_{im}$ is the penalty function regularization parameter. The PSO fitness function can therefore be expressed as follows:
\begin{equation}
\small
F(\boldsymbol{X}) = f(\boldsymbol{X}) + \sum_{m \in \mathcal{M}}\sum_{i \in \mathcal{I}}P_{im}(x_m, z_m, \epsilon_{m}). 
\end{equation}


\section{Selected Numerical Results}
\label{results_sec}
In this section, we discuss the logic behind model parameters selected for the initial model run, and discuss the meaning of the results. After, we perform a brief sensitivity analysis in order to test the robustness of the model across varying circumstances. 
Table~\ref{modelparams} summarizes the values chosen for the initial run, which were based on the physical aspects of the Velodyne radar~\cite{Gunnam2018}, along with an ideal representation of a typical three-lane, one-way urban street, with lane widths of 5 m (total roadway width of 15 m), and ideal parameters for fiber-optic communications between the ELiD's and the backhaul network.

\begin{table}[t]
\scriptsize
\centering
\label{paramTable}
\caption{Initial Model Run Parameters}
\addtolength{\tabcolsep}{-4pt}\begin{tabular}{|c|c|c||c|c|c||c|c|c|}
\hline
\textbf{Parameter} &
\textbf{Value} &
\textbf{Unit} &
\textbf{Parameter} &
\textbf{Value} &
\textbf{Unit} &
\textbf{Parameter} &
\textbf{Value} &
\textbf{Unit} \\
\hline
$\lambda$ &
0.25 &
n/a &
$\eta$ &
1 &
n/a &
$D_{road}$ &
1 &
km \\
\hline
$\bar{B}$ &
10 &
GB/s &
$f_{scan}$ &
30 &
Hz &
$d$ &
5 &
n/a\\
\hline
$P_{comm}$ &
5 &
W &
$R_{comm}$ &
1 &
GB/s&
$P_{rad}$ &
10 &
W  \\
\hline
$\bar{E}$ &
20 &
W &
$y_{min}$ &
5 &
m &
$y_{max}$ &
20 &
m\\
\hline
$|\mathcal{M}|$ &
20 &
n/a &
$H_{cov}$ &
2 &
m &
$\theta$ &
120 &
deg. \\
\hline
$\phi$ &
35 &
deg &
$z_{min}$ &
15 &
m &
$z_{max}$ &
50 &
m\\
\hline
$\alpha$ &
1 &
n/a &
$\beta_{g}$ &
2 &
n/a &
$\beta_{p}$ &
2 &
n/a\\
\hline
$\delta$ &
$10^{-1}$ &
deg &
$\gamma$ &
$10^{-1}$ &
m &
$\xi$ &
$10^{-4}$ &
m\\
\hline
$|S|$ &
100 &
n/a &
$\rho_{im}, \forall i, m$ &
$1$ &
n/a &
$t_{max}$ &
500 &
iterations\\
\hline
\hline
\textbf{Parameter} &
\multicolumn{7}{|c|}{\textbf{Values}} &
\textbf{Unit}\\
\hline
$Q_a$ &
\multicolumn{7}{|c|}{0.06, 0.15, 0.3, 0.7, 0.86, 0.94, 1} &
km \\
\hline
$\Delta_a$ &
\multicolumn{7}{|c|}{1, 0.9, 0.8, 0.77, 0.8, 0.9, 1} &
n/a\\
\hline
\end{tabular}
\label{modelparams}
\end{table}

\begin{figure*}[t!]
\centering
\includegraphics[width=18cm]{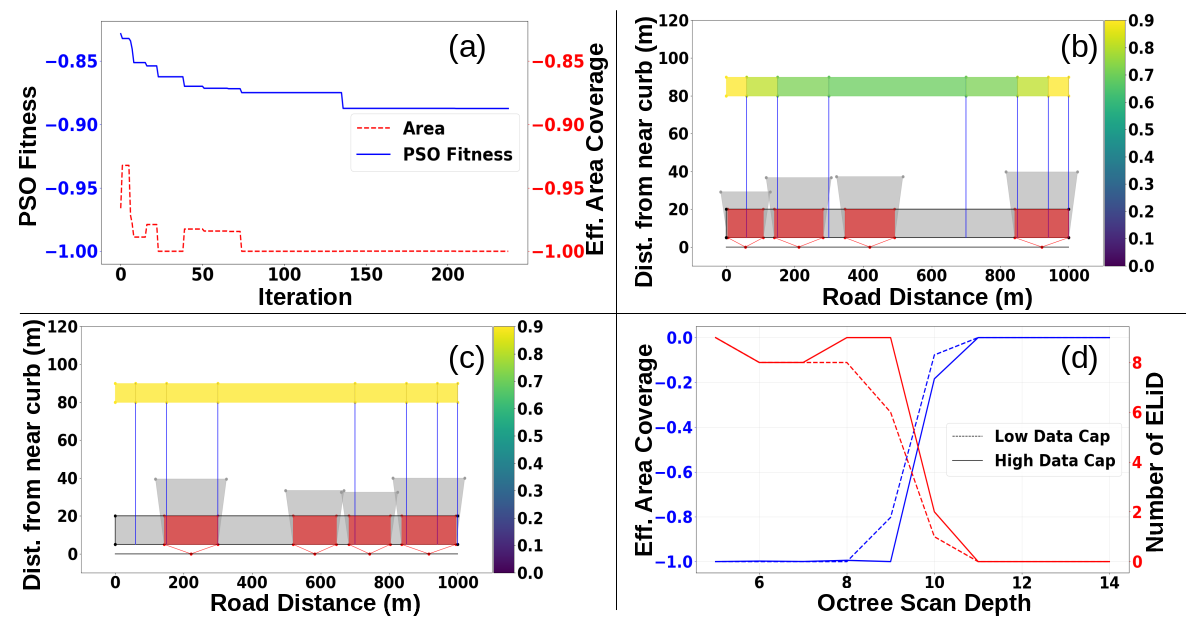}
\caption{(a) Initial model run cost function convergence. The solid blue line corresponds to the PSO fitness function, and the dashed red line corresponds to the effective area coverage ($-A^{eff}_{cov}$). (b) and (c) A comparison of placement with more restrictive parameters than the initial run, where (b) considers roadway areas with varying importance metrics, and (c) considers roadway segments with equal importance metrics; placement is affected by the importance metric. In this case, the segments of roardway near 0 and $D_{road}$ are at intersections. The red squares represent the effective coverage area of the ELiD ($A^{Rectangle}_m$). The gray trapezoids represent the total area ($A^{Total}_m$). The blue vertical lines represent the divisions of the roads ($Q_a$) based on their importance metrics ($\Delta_a$). The roadway is represented as a black square ranging in the x-axis from $[0, D_{road}]$, and in the y-axis from $[y_{min}, y_{max}]$. (d) The results from the sensitivity analysis - dashed lines for $\bar{B} = 5$ GB/s, and solid lines for $\bar{B} = 10$ GB/s.}
\label{results}
\end{figure*}

Fig.~\ref{results}(a) illustrates the initial run convergence. We observe that the PSO solver capably converges to an efficient local optimum for this cost function where the entire roadway is covered. There is a minimal amount of excess overlap between the ELiD coverage zones, and there are no constraint violations. Fig.~\ref{results}(b) and (c) illustrate the effects of the importance metric on placement - showing that the solver can adjust placement based on the necessity of certain stretches of road receiving coverage. For a fair comparison, we set a higher scan depth $d=9$ and lower data throughput capacity $\bar{B} = 3.6$ while keeping the other initial model run parameters constant to show sparse resource allocation based on the "importance". In the case of non-optimal placement of the ELiD's (i.e. placed at $x_m = 125, 375, 625, 875$, and all at $z_m=40$), $A_{cov}^{eff} = 0.552$, where the coverage efficiency for PSO-optimized placement (Fig.~\ref{results}(b)) is $A_{cov}^{eff} = 0.570$. 

We also perform a sensitivity analysis on the octree scan depth parameter $d$, as it directly affects both the energy constraint and throughput capacity constraints (i.e. an increase in $d$ leads to increased energy use and data transmission). While varying the $d$ parameter, we also test performance at different throughput capacities $\bar{B}$, to visualize how the varying data generation parameter has effects with more or less generous throughput capacities. We notice, from Fig.~\ref{results}(d), that increasing the depth parameter leads to less effective coverage, as the increased data generated by each ELiD is more taxing on the system. We also observe that a lower throughput cap leads to this effect getting magnified. The reduction in coverage as $d$ increases shows that the PSO solver finds feasible solutions -- $A^{eff}_{cov} = 0$ if the placement of even a single ELiD violates constraints.


\section{Conclusion}
\label{conclusion}
In this paper, we optimized the placement of ELiD lamps to maximize the effective urban roadway coverage using an evolutionary algorithm, i.e., PSO. It is shown that coverage efficiency is constrained by physical restrictions such as energy consumption and bandwidth capacity. Covering urban roadways with ELiD technology remains a challenging task especially when budget limitations are also considered.


\bibliographystyle{ieeetr}
\bibliography{references}

\end{document}